\documentclass{article}
\usepackage{spconf,amsmath,graphicx}
\usepackage{url}
\usepackage{booktabs}
\usepackage{amssymb}

\newcommand{\eg}{\textit{e.g.}}
\newcommand{\ie}{\textit{i.e.}}

\usepackage{xcolor}
\usepackage{bbding}
\usepackage{pifont}

\usepackage{cleveref}
\crefname{appendix}{App.\negthinspace\,}{App.\negthinspace\,}
\crefname{chapter}{Chap.\negthinspace\,}{Chap.\negthinspace\,}
\crefname{equation}{Eq.\negthinspace\,}{Eq.\negthinspace\,}
\crefname{algorithm}{Alg.\negthinspace\,}{Alg.\negthinspace\,}
\crefname{section}{Sec.\negthinspace\,}{Sec.\negthinspace\,}
\crefname{subsection}{Sec.\negthinspace\,}{Sec.\negthinspace\,}
\crefname{subsubsection}{Sec.\negthinspace\,}{Sec.\negthinspace\,}
\crefname{figure}{Fig.\negthinspace\,}{Fig.\negthinspace\,}
\crefname{table}{Tab.\negthinspace\,}{Tab.\negthinspace\,}
\crefname{subfigure}{Fig.\negthinspace\,}{Fig.\negthinspace\,}
\crefname{subsubfigure}{Fig.\negthinspace\,}{Fig.\negthinspace\,}
\crefname{lstlisting}{Lst.\negthinspace\,}{Lst.\negthinspace\,}

\title{CellCycleGAN: Spatiotemporal Microscopy Image Synthesis of Cell Populations using Statistical Shape Models and Conditional GANs}
\name{{\parbox[c]{\textwidth}{\centering Dennis B\"ahr$^{1,\#}$, Dennis Eschweiler$^{1}$, Anuk Bhattacharyya$^{1}$, Daniel Moreno-Andrés$^{2}$, Wolfram Antonin$^{2}$, Johannes Stegmaier$^{1,\#,*}$\thanks{$^{\#}$Authors contributed equally, $^*$Correspondence: \texttt{johannes.stegmaier@lfb.rwth-aachen.de}}}}}

\address{
\footnotesize $^{1}$Institute of Imaging and Computer Vision, RWTH Aachen University, Aachen, Germany\\
\footnotesize $^{2}$Institute of Biochemistry and Molecular Cell Biology, Medical School, RWTH Aachen University, Aachen, Germany\\
}

\begin{document}
%\ninept
%
\maketitle
\begin{abstract}
    Automatic analysis of spatio-temporal microscopy images is inevitable for state-of-the-art research in the life sciences. Recent developments in deep learning provide powerful tools for automatic analyses of such image data, but heavily depend on the amount and quality of provided training data to perform well. To this end, we developed a new method for realistic generation of synthetic 2D+t microscopy image data of fluorescently labeled cellular nuclei. The method combines spatiotemporal statistical shape models of different cell cycle stages with a conditional GAN to generate time series of cell populations and provides instance-level control of cell cycle stage and the fluorescence intensity of generated cells. We show the effect of the GAN conditioning and create a set of synthetic images that can be readily used for training and benchmarking of cell segmentation and tracking approaches.
\end{abstract}
\section{Introduction}
\label{sec:intro}

Automatic segmentation and tracking is indispensable to be able to keep pace with the amount of data acquired in high-throughput microscopy experiments. Deep learning has recently largely transformed the field of computer vision and provides excellent results on various imaging domains \cite{Meijering2020}. However, this only holds true if the models are trained with sufficient training data. Despite community efforts like the Broad Bioimage Benchmark Collection and the Cell Tracking Challenge \cite{Ulman17}, manually annotated training data sets for the field of bioimage analysis are still rare. Data synthesis has thus been of high interest, \eg, to augment manually annotated training data sets or to provide data sets for algorithm benchmarking \cite{Ulman17} and image restoration \cite{Kozubek2020}.

Cell image synthesis is often considered a three-step approach with the initial generation of cell phantoms, texture synthesis and a final simulation of the imaging system \cite{Svoboda2016}. The morphology can be modeled, \eg, using prior knowledge-based deformations of basic shapes \cite{Svoboda2016}, statistical shape models \cite{Heimann09}, spherical harmonics \cite{Ruan2019} or using shape spaces derived from diffeomorphic measurements \cite{Peng2009}.
As a next step, generated phantoms need to be translated to realistically looking images, which can be obtained either classically by mathematic description of the texture synthesis, by modeling protein distributions in sub-cellular components or by a transfer of real textures to the simulated objects \cite{Kozubek2020}. With the advent of generative adversarial networks (GANs), deep convolutional neural networks demonstrated to excel at realistic image data generation as well \cite{Isola17, Zhu2017}. Extensions of the GAN framework with conditional labels allow generating realistic images that reflect semantic properties provided to the generator \cite{Isola17}. Recently, these methods were also used for generating biological images like multi-channel data of human cultured cells \cite{Goldsborough2017}, protein localization in different cell cycle stages \cite{Osokin2017} or entire tissues \cite{Eschweiler19_SASHIMI}. In addition to generation of realistic textures, GANs have also been successfully used to mimic the shape of cells in 3D \cite{Wiesner2019}. After objects have been placed in a virtual image, the simulations are usually finalized by adding artificial disruptions like dark current, photon shot noise, sensor readout noise and a point spread function \cite{Svoboda2016,Stegmaier16_EmbryomicsBenchmark} or using more elaborate physically motivated wave-optical simulation approaches \cite{Weigert18b}.

While GANs are certainly the state-of-the-art with respect to achievable realism of synthetic images, their application in biomedical imaging was mainly style transfer and the generation of static time points. Thus, the synthesis of temporally changing morphologies and sufficient control over the generation process are still missing. We address these limtations in the current work and our main contributions are (1) time-resolved statistical shape models to create masks that mimic the cell morphology of different cell cycle stages, (2) a GAN for image synthesis including cell cycle stage and intensity contitioning and (3) a framework to generate realistic 2D+t image sequences with varying amounts of cells, dynamically moving objects and randomized cell behavior. We demonstrate the functionality of the GAN conditioning and generate a set of realistic synthetic image data that can be used to train and benchmark cell segmentation and tracking algorithms.

\section{CellCycleGAN}

\subsection{Stage Sequence Generation using a Graphical Model}
\begin{figure}[h]
    \includegraphics[width=\columnwidth]{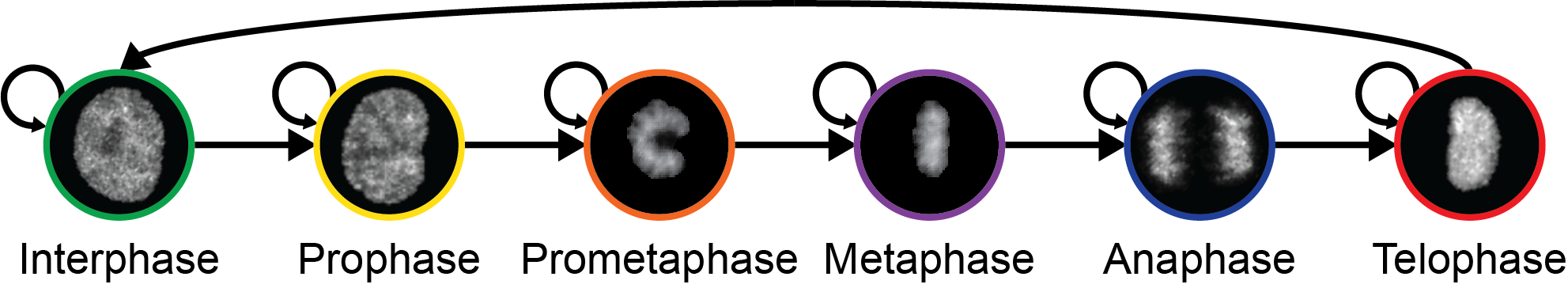}
    %\vspace{-0.7cm}
    \caption{Graphical model to generate synthetic stage sequences. Colors indicate different stages and are encoded as labels $1 - 6$ from left to right for the GAN training. Transition probabilities are estimated from annotated training data including constraints on minimum and maximum duration.}
    \vspace{-0.3cm}
    \label{fig:Figure1}
\end{figure}

The first step of the proposed pipeline is the generation of an artificial sequence of states of a mitotic cycle. Based on the chromatin morphology, the cell cycle stages are typically grouped into interphase, prophase, prometaphase, metaphase, anaphase and telophase that ends in another interphasic stage  \cite{Held10}. Analogous to previous work \cite{Held10, Zhong12}, we model this process using a graphical model depicted in \cref{fig:Figure1}. The model comprises six nodes, \ie, one node for each cell cycle stage. To allow remaining in a particular stage for multiple frames, each of the nodes has a connection to itself. We start in a randomly selected initial stage and sample the next stages from the graphical model. The transition probabilities were estimated from the ground truth data by counting the frequency of particular stage transitions. Moreover, we determined the minimum and maximum durations of each phase and use these as hard constraints during the sampling, \ie, the model resides in the current stage if the minimum duration was not yet met and directly jumps to the next stage if the maximum duration of the stage is exceeded. 

\subsection{Temporally Variant Morphology Modeling using Multiple Statistical Shape Models}
We rely on a classical approach for generating various cell morphologies using statistical shape models derived from a small set of manually labeled images \cite{Heimann09, Zhong12} . In brief, each cell shape is characterized by a set of evenly distributed landmarks on its boundary. To be able to compare shapes that are oriented differently, all segmentation masks are centered at the origin and rotated such that the major axis points upwards. We sample landmarks on the boundaries in angular steps of $6^\circ$ and thus end up with $60$ landmarks per segmentation mask. Due to the regular sampling scheme, statistics of corresponding landmarks can be computed, \ie, we compute the mean shape and a shape covariance matrix of the landmarks. As the different stages observed during a mitotic cycle exhibit different morphologies, we model each of the phases with a separate statistical shape model indicated by subscript $s \in \lbrace 1, ... ,6 \rbrace $, where each of the shape models is computed from $N_s$ ground truth shapes. The mean shape $\bar{\mathbf{x}}_s$ of shape model $s$ is computed on vectorized landmark coordinates $\mathbf{x}_{si}, i \in \lbrace 1, ..., N_s \rbrace$ as:
\begin{equation}
    \bar{\mathbf{x}}_s = \frac{1}{N_s} \sum_{i=1}^{N_s} \mathbf{x}_{si},
\end{equation}
and the shape covariance matrix $\mathbf{\Sigma}_s$  given by:
\begin{equation}
    \mathbf{\Sigma}_s = \frac{1}{N_s-1} \sum_{i=1}^{N_s} (\mathbf{x}_{si} - \bar{\mathbf{x}}_s)(\mathbf{x}_{si} - \bar{\mathbf{x}}_s)^T.
\end{equation}
Determining the eigenvectors $\mathbf{e}_{si}$ and eigenvalues $\lambda_{si}$ of these shape covariance matrices yields the principal directions of variation of all landmarks ordered in descending order via the eigenvalues $\lambda_{si}$. It is thus possible to approximate the shape of a cell $\tilde{\mathbf{x}}_{s}$ at a particular stage $s$ using the mean shape and a linear combination of the $N_e$ eigenvectors or a subset thereof:
\begin{equation}
    \tilde{\mathbf{x}}_s = \bar{\mathbf{x}}_s + \sum_{i=1}^{N_e} b_{si} \mathbf{e}_{si},
\end{equation}
with scalar weighting factors $b_{si}$. Artificial new shapes can be generated by sampling an $N_e$-dimensional random vector $\mathbf{b}_s$ from a Gaussian distribution $\mathcal{N}(0,1)$ and by additionally scaling it relative to the associated eigenvalues $\sqrt{\lambda_{si}}$.

To model a sequence of different stages, we employ all identified shape models and perform a weighting of the shape models. First, all stage transitions in a sequence are determined and we temporally place Gaussian kernels centered above each transition point with $\sigma_s$ set to the frames per stage:  
\begin{equation}
    w_s(t) = \frac{1}{\sqrt{2\pi\sigma_s^2}} e^{-\frac{(t - \mu)^2}{2\sigma_s^2}}.
\end{equation}
A new shape $\tilde{\mathbf{x}}(t)$ at a time $t$ can be assembled as a weighted sum of the shape models of all stages, where a sequence of states with the same label uses the same random vector $\mathbf{b}_{si}$:
\begin{equation}
\tilde{\mathbf{x}}(t) = \sum_{s=1}^{6} \frac{w_s(t)}{\sum_{k=1}^{6} w_k(t)} \left( \bar{\mathbf{x}}_s + \sum_{i=1}^{N_e} \epsilon \cdot \sqrt{\lambda_{si}} \cdot (\mathbf{b}_{si} \circ \mathbf{e}_{si})  \right).
\end{equation}
 Each eigenvector is additionally scaled with a small normally distributed random factor $\epsilon \sim \mathcal{N}(0,0.1)$, such that multiple instances within a stage are slightly varying. The random factors were empirically determined and yielded sufficiently different shapes while avoiding to sample unrealistic shapes.

\subsection{Conditional Generative Adversarial Networks for Stage-Specific Texture Synthesis}

To transform the single-cell mask time series to realistically looking images of the respective stages, we employ a conditional generative adversarial network inspired by Pix2Pix and the cLR-GAN approaches \cite{Isola17, Zhu2017}. We condition the generator using three channel input images, where the first channel is a discrete mask image with zero values in background regions and with the respective cell cycle stage encoded in the label intensity of the foreground region. Moreover, an intensity conditioning is performed by supplying a second channel with the desired mean intensity encoded in the mask. To randomize the generation procedure, we add a third channel with uniformly distributed noise from the interval $\left[0,1\right]$. We use a small U-Net-like architecture with four levels as the generator networks \cite{Ronneberger15}. Each level consists of two convolutional layers with kernel size $3 \times 3$, stride $1$, instance normalization and ReLU activation. We start with $16$ feature maps and double the depth after each downsampling step with a maximum depth of 128 feature maps. Upsampling is performed with transposed convolution with a kernel size of $3 \times 3$, stride $2$ and includes skip-connections from the downsampling path. Finally, a $1 \times 1$ convolutional layer maps to a single output channel with a sigmoid activation that should represent the generated raw image. 
We attach a second encoder-decoder network to this output of the first part of the network. The second encoder-decoder network has the same architecture as described above except that it has two output channels, which intend to reconstruct the conditional labels for stage and intensity conditioning. Adding the label reconstruction path is intended to force the network to make use of the conditional labels and is similar to one direction of the cycle consistency loss employed in the CycleGAN architecture. We note that adding a cycle consistency loss in the other direction would not make sense due to the one-to-many mapping of a mask to multiple raw images. 

The discriminator uses the common PatchGAN architecture and is trained with both generated and real image patches including the corresponding channels for stage and intensity conditioning. Thus, the discriminator should learn to separate real from fake images and additionally to distinguish different cell cycle stages, by detecting a mismatch between the conditioned cell appearance and the corresponding raw image. We use binary cross-entropy as the adversarial loss of the discriminator and an L1 loss term to encourage a good reconstruction of the contitional labels. The model is implemented in PyTorch and trained for 1000 epochs using Adam with a constant learning rate of $0.001$.

\subsection{Putting it all Together}

Finally, we want to generate images of a whole cell population rather than single-cell videos. Thus, we create simulations with a desired number of objects initialized at uniformly distributed random start positions. As the first step, we sample sequences of stages for each of the objects from the graphical model and use these stages to generate artificial masks using the statistical shape models for a predefined number of frames. As soon as a metaphase to anaphase transition is sampled, we introduce a split event and initialize two daughter cells. To ensure smooth stage transitions, the shape models of both daughter cells are initialized with the same shape model of the late anaphase stage after the segregation of mother cell chromatin. From there on, subsequent cell cycle stages of each daughter are sampled independently and thus develop differently. For each stage, we estimated the average intensity and an average standard deviation from real microscopy images. The intensity conditioning is determined by this stage-dependent mean intensity and a randomly drawn offset that is similarly applied to the entire lineage. Thus, a lineage that is more dim in the beginning of the time series will remain dim even after division events. The shape, stage and intensity properties of each cell are then used to generate artificial microscopy images using the GAN. 

In addition to morphology and appearance, the dynamic movement of the cells is important for a realistic impression. We assume a Brownian motion model and displace each object in each frame by a two-dimensional random vector sampled from a normal distribution $\mathcal{N}(0,2)$ and randomly rotate each object relative to its predecessor with an angle drawn from another normal distribution with $\mathcal{N}(0,1)$. Upon the division of a cell, the daughter cells are displaced perpendicular to the major axis of the mother cell, which is a typical behavior observed in real cells upon division. In order to avoid overlapping cells after multiple rounds of cell divisions, a repulsive force as described by Macklin et al. \cite{Macklin12} is applied. As estimation of the $R_M$ and $R_N$ parameters, we use the major and the minor axis length of each object. Thus, neighbors that are residing nearby the reference nucleus are slightly pushed apart. The final images are subsequently processed with a classical acquisition simulation involving a fixed offset for the dark frame, Gaussian smoothing as a point spread function approximation ($\sigma=1$), Poisson shot noise and Gaussian noise ($\mu=0, \sigma=0.001$) for sensor noise \cite{Stegmaier16_EmbryomicsBenchmark}.
\begin{figure}[h!]
    \includegraphics[width=\columnwidth]{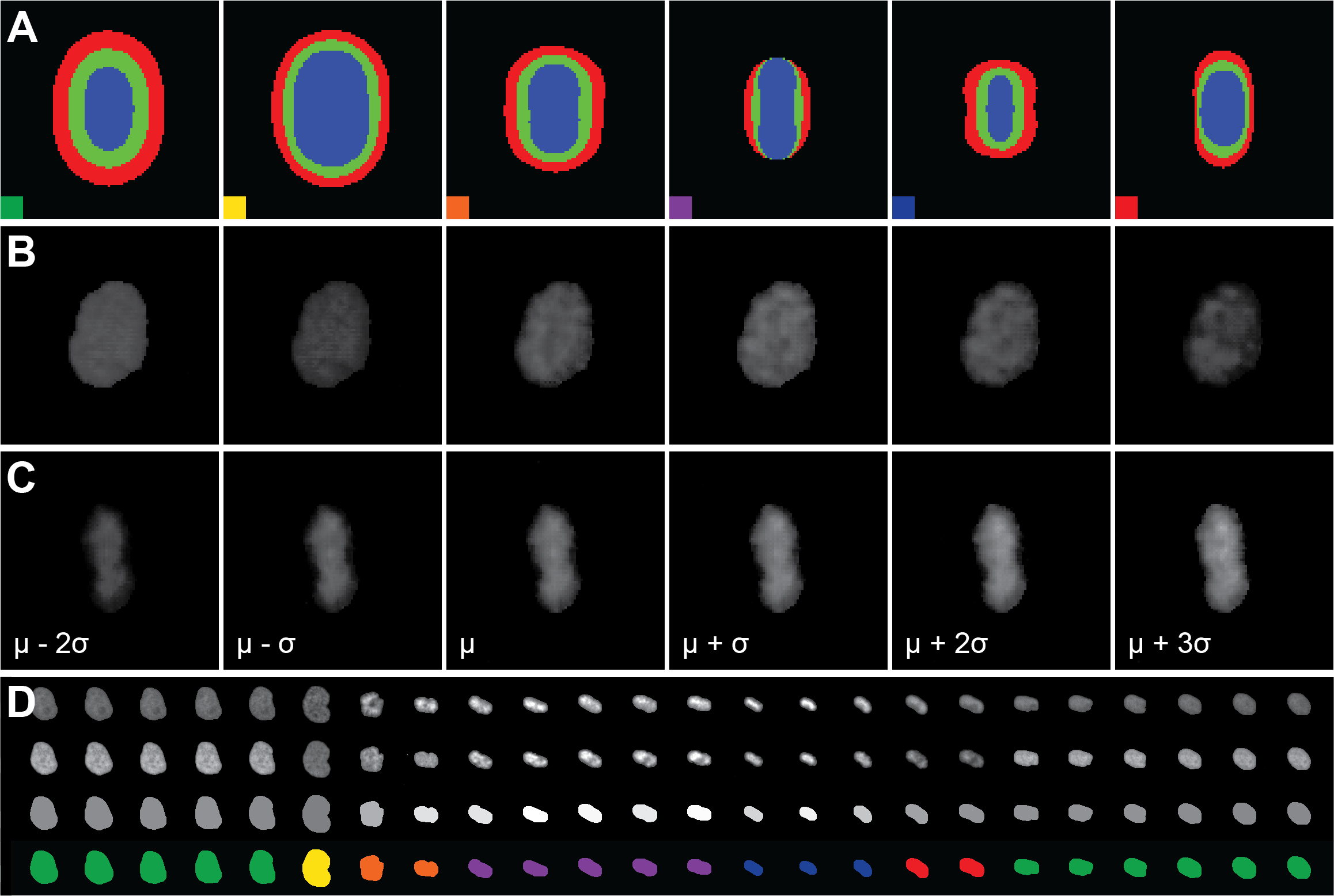}
    %\vspace{-0.7cm}
    \caption{Qualitative analysis of the different simulation steps. (A) Mean shapes (green) $\pm$ the first eigenvector (red and blue) of the statistical shape models. (B) GAN-generated images for varying stage conditioning with fixed shape and intensity conditioning. (C) GAN-generated images for varying intensity conditionings with fixed shape and stage. Intensity values are based on the mean intensity and standard deviation observed in real microscopy images. (D) From top to bottom: real images, GAN-generated images using real masks for comparability, intensity conditioning and stage conditioning. Intensity adjusted for better visibility; stages as in \cref{fig:Figure1}.}
    \vspace{-0.4cm}
    \label{fig:Figure2}
    \end{figure}

\section{Generating Synthetic 2D+t Microscopy Images of Fluorescent Nuclei}
We tested our approach on a publicly available data set where the different mitotic stages were manually labeled \cite{Zhong12}. The data set consists of time-resolved crops of single cells from seven experiments that were imaged with a widefield microscope (10$\times$, $0.5$ NA) with a total of $304$ cells that were tracked over $40$ frames using the CellCognition software \cite{Held10}. The available manual annotations comprise stage labels for each of the image frames, grouping the frames into the six distinct phases interphase, prophase, prometaphase, metaphase, anaphase and telophase (\cref{fig:Figure1}). To facilitate the subsequent processing, each image was scaled to $96 \times 96$ pixels and rotated such that the major axis of the object coincided with the y-axis. We segmented the snippets using a simple threshold combined with a seeded watershed on the inverted distance transform of the binary image.

\begin{figure}[h]
    \includegraphics[width=\columnwidth]{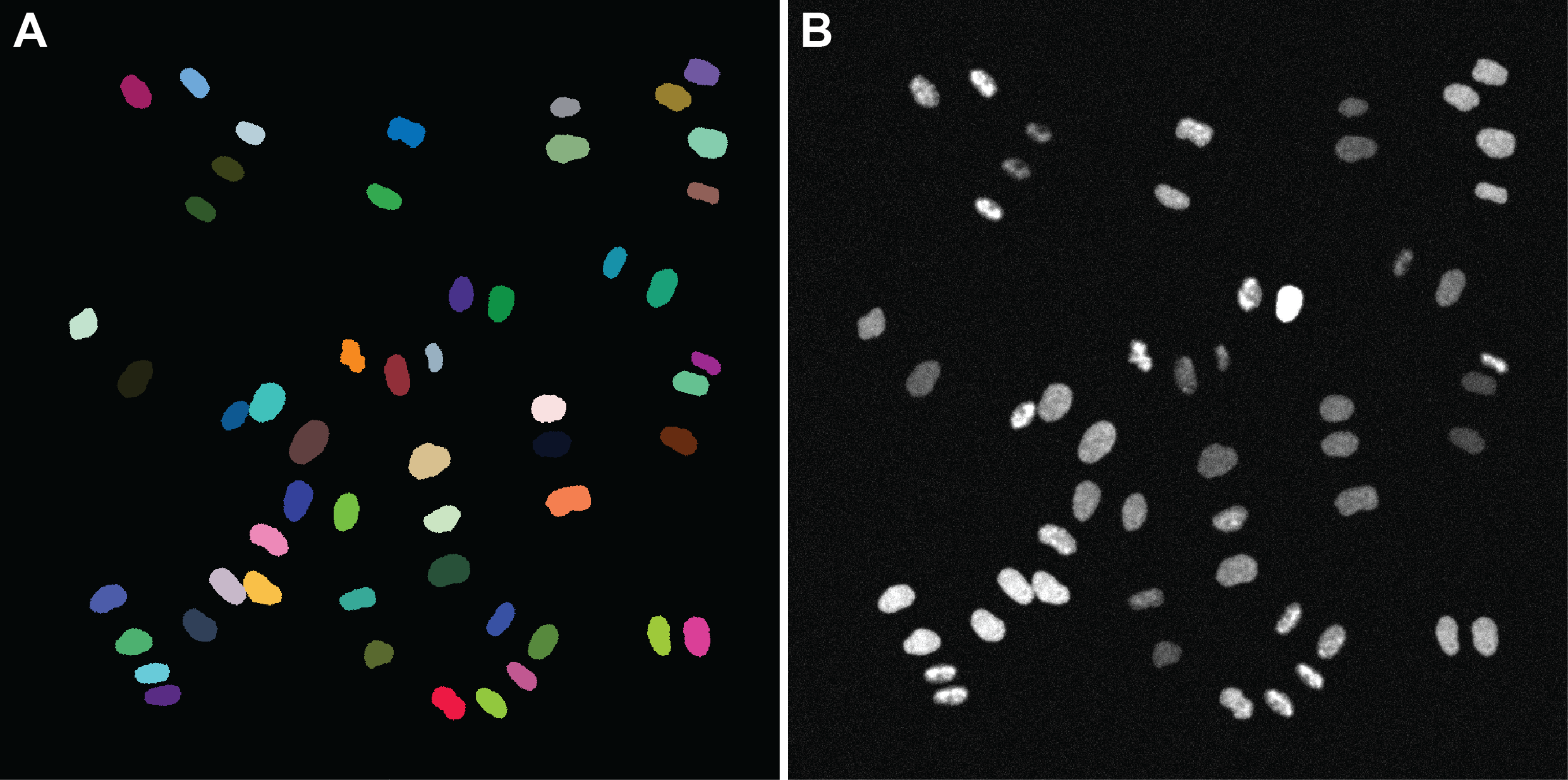}
    %\vspace{-0.7cm}
    \caption{Spatiotemporal simulation of a cell population. (A) Randomly colored instances that were used for the GAN-based raw image synthesis. (B) Generated raw image assembled from single-cell image patches including postprocessing.}
    \vspace{-0.3cm}
    \label{fig:Figure3}
\end{figure}

The different shape models that were computed separately for each of the mitotic stages exhibit clear size and shape variations that resemble the morphology of real cells during these stages (\cref{fig:Figure2}A). Due to the random parameterization of the shape models, each generated cell will slightly differ and the deviation from the mean shape can be controlled by weighting the random contributions of the eigenvectors of the shape covariance matrix. Next, we investigated the effect of the two input channels used for the conditioning of the generator network. While keeping the intensity conditioning and the generated shape fixed, \cref{fig:Figure2}B showcases different textures that were generated by varying the stage conditioning. The intensity gradient depicted in \cref{fig:Figure2}C illustrates the effect of successively increasing intensity conditioning while keeping shape and stage conditioning fixed. A comparison of a sequence of real images to synthetic single-cell snippets including the used conditional labels is depicted in (\cref{fig:Figure2}D). A full frame comprising stage-dependent variation of the statistical shape models in combination with the corresponding GAN conditioning for multiple interacting cells with different shapes, stages and appearances is shown in \cref{fig:Figure3}. In addition to the synthetic images, a comprehensive ground truth including instance segmentation, cell lineages and cell stages is available that can be used for supervised training of dedicated segmentation or tracking methods. Moreover, the postprocessing steps can be used to generate artificial images of varying difficulty levels, \eg, to train supervised models with very difficult images and thus to potentially make them more robust for the variability observed in real applications. We quantified the performance of the GAN-based synthesis with the Fréchet Inception Distance (FID, implementation by M. Seitzer) \cite{Heusel2017} and obtained values of $66.6$ (real vs. generated) and $146.6$ (real vs. intensity cond. mask), indicating a high similarity of the distribution learned by the GAN and the real underlying microscopy data. Source code, generated image data sets and example videos can be obtained from \url{https://github.com/stegmaierj/CellCycleGAN}.

\section{Conclusions}
In this contribution we present a new framework for the synthesis of artificial 2D+t microscopy images of fluorescently labeled cell nuclei. The simulation involves temporally changing statistical shape models to generate synthetic shapes of various cell morphologies observed throughout the mitotic cycle. To generate realistically looking synthetic objects, we train a conditional GAN with the possibility to control both stage-dependent appearance and the intensity of the synthetic images. Initially, we create single-cell video snippets and assemble larger images including a simple movement simulation, repulsive object interactions and postprocessing.

While the single-cell snippets nicely resemble the appearance of real microscopy images, the assembled multi-cell images still look somewhat artificial. This could be compensated, \eg, by exploiting the fully-convolutional GAN architecture for generating an entire multi-cell frame, using an additional unpaired style transfer as a postprocessing step or by incorporating spatiotemporal constraints to further push the level of realism. Very short cell cycle stages like the prophase are underrepresented in the training data set and are not yet realistic enough. We would thus have to enlarge the training database or rely on balanced sampling strategies instead. Future work includes modeling more complex cell shapes and the extension to 3D+t, where annotated training data is even harder to get. Furthermore, the investigation of unsupervised and continuous identification of the stage transitions would get rid of the required manually identified stages, \eg, using combinations of GANs and LSTM networks.

\clearpage
\section{Compliance with Ethical Standards}
The work deals with synthesis of artificial image data for which no ethical approval was required. 

\section{Acknowledgements}
This work was funded by the German Research Foundation DFG with the grants STE2802/2-1 (DE), ME3737/3-1 (AB) and by the Excellence Initiative of the German federal and state governments as an RWTH StartUp project (AB).

\bibliographystyle{IEEEbib}
%\bibliographystyle{diss}
%\bibliography{../../../Literature/Bibliography}

\end{document}